 \newfont{\goth}{cmbxti10 scaled\magstep1}
 \newfont{\gothi}{cmbxti10}

 \newcommand{\smin}{\,\raisebox{0.06em}{${\scriptstyle \in}$}\,}

 \newcommand{\Integer}{\:\mbox{\sf Z} \hspace{-0.82em} \mbox{\sf Z}\,}

 \newcommand{\gotg}{\mbox{\goth g}}

\def\maketitle{
  \par
  \begingroup
    \def\thefootnote{\fnsymbol{footnote}}
    \def\@makefnmark{\hbox to 0pt{$^{\@thefnmark}$\hss}}
    \if@twocolumn
      \twocolumn[\@maketitle]
    \else
      \newpage
      \global\@topnum\z@
      \@maketitle
    \fi
    \thispagestyle{empty}                                           
    \@thanks
  \endgroup
  \setcounter{footnote}{0}
  \setcounter{page}{0}                                              
  \let\maketitle\relax
  \let\@maketitle\relax
  \gdef\@thanks{}
  \gdef\@author{}
  \gdef\@title{}
  \let\thanks\relax }

\def\ps@myfootings{
  \def\@oddfoot{\hbox{}\sl\rightmark \hfil\rm\thepage}
  \def\@evenfoot{\rm \thepage\hfil\sl\leftmark\hbox{}}
  \def\@oddhead{}
  \def\@evenhead{}
  \def\sectionmark##1{}
  \def\subsectionmark##1{} }

\def\eqnumpersection{ \@addtoreset{equation}{section}
                      \def\theequation{\thesection.\arabic{equation}} }
\@definecounter{eqnlist}

\newcount\@eqnlcnt
\newcount\@eqnlpen

\def\eqnlist{
  \setcounter{eqnlist}{1}
  \let\@currentlabel=\theeqnlist
  \global\@eqnswtrue \global\@eqnlcnt\z@
  \tabskip \@centering
  \let\\=\@eqnlcr
  $$\halign to \displaywidth\bgroup\@eqnlsel\hskip\@centering
  $\displaystyle\tabskip\z@{##}$&\global\@eqnlcnt\@ne
  \hskip 2\arraycolsep \hfil${##}$\hfil
  &\global\@eqnlcnt\tw@ \hskip 2\arraycolsep
$\displaystyle\tabskip\z@{##}$\hfil
  \tabskip\@centering&\llap{##}\tabskip\z@\cr }

\def\endeqnlist{ \@@eqnlcr\egroup $$\global\@ignoretrue }

\let\@eqnlsel=\relax

\def\@eqnlcr{ {\ifnum0=`}\fi
              \@ifstar{\global\@eqnlpen\@M\@yeqnlcr}
                      {\global\@eqnlpen\interdisplaylinepenalty \@yeqnlcr} }

\def\@yeqnlcr{ \@ifnextchar [{\@xeqnlcr}{\@xeqnlcr[\z@]} }

\def\@xeqnlcr[#1]{ \ifnum0=`{\fi}\@@eqnlcr
                   \noalign{\penalty\@eqnlpen\vskip\jot\vskip #1\relax} }

\def\@@eqnlcr{ \let\@tempa\relax
               \ifcase\@eqnlcnt \def\@tempa{& & &}\or \def\@tempa{& &}
               \else \def\@tempa{&}\fi
               \@tempa
               \if@eqnsw \@eqnlnum \stepcounter{eqnlist} \fi
               \global\@eqnswtrue \global\@eqnlcnt\z@\cr }

\def\@eqnlnum{\hbox to .01pt{}\rlap{\rm \hskip -\displaywidth ~~~\theeqnlist)}
}

\def\theeqnlist{\roman{eqnlist}}

\documentstyle[12pt,uartopt]{article}

\input{fordef.tex}

\begin{document}

\renewcommand{\thefootnote}{\fnsymbol{footnote}}

\title{Current Algebra of WZNW Models at and away from Criticality}

\author{E.\ Abdalla$\,^1$ \footnotemark[1]~~~~and~~
        M.\ Forger$\,^2$ \footnotemark[2]}
 \footnotetext[1]{Work partially supported by CNPq}
 \footnotetext[2]{Work partially supported by FAPESP}

 \date{\normalsize
       $^1\,$ Instituto de F\'{\i}sica, Universidade de S\~ao Paulo, \\
       Cx.\ Postal 20516, BR-01498 S\~ao Paulo SP / Brazil \\
       e-mail: eabdalla @ uspif.if.usp.br \\[0.4cm]
       $^2\,$ Fakult\"at f\"ur Physik der Universit\"at Freiburg, \\
       Hermann-Herder-Str. 3, W-7800 Freiburg / FRG \\
       e-mail: forger @ ibm.ruf.uni-freiburg.de}

\maketitle
\thispagestyle{empty}
\begin{abstract}
\noindent
We derive the current algebra of principal chiral models with a Wess-Zumino
term. At the critical coupling where the model becomes conformally invariant
(Wess-Zumino-Novikov-Witten theory), this algebra reduces to two commuting
Kac-Moody algebras, while in the limit where the coupling constant is taken
to zero (ordinary chiral model), we recover the current algebra of that model.
\linebreak
In this way, the latter is explicitly realized as a deformation of the former,
with the coupling constant as the deformation parameter.
\end{abstract}

\vfill

 \begin{flushright}
 \parbox{12em}
 { \begin{center}
 University of Freiburg \\
 THEP 92/10 \\
 Universidade de S\~ao Paulo \\
 IFUSP/P- 972\\
 March 1992
 \end{center} }
 \end{flushright}

\newpage
\renewcommand{\thefootnote}{\arabic{footnote}}

\setcounter{page}{1}

Algebraic methods have during the last few years been used extensively to
unravel the dynamical structure of two-dimensional field theoretical models,
especially integrable models and conformal field theories. In the case of
conformally invariant models, a purely algebraic approach (using, e.g., the
representation theory of the Virasoro algebra and of Kac-Moody algebras,
together with that of braid groups) may even lead to a complete solution
of the theory.

In the present paper, we shall investigate the current algebra for the
principal chiral model with a Wess-Zumino term. This model \cite{W}
contains a free coupling constant $\lambda$ and, for special values of
$\lambda$, contains the conformally invariant WZNW model as well as the
ordinary chiral model. Therefore, the current algebra derived below is a
common generalization of the current algebras for these two special cases.
Now for the WZNW model, the current algebra is well-known to consist of
two commuting Kac-Moody algebras, while for the ordinary chiral model, it
is a new kind of algebra which -- although having been known in part for
some time (cf., e.g., \cite[pp.\ 323/324]{FaTa}) -- has only recently been
specified completely \cite{FLS} and whose mathematical structure is still
far from being understood. The hope is that through the ``interpolating''
current algebra derived here, the well-developed theory of Kac-Moody algebras
may shed some light on the current algebra for integrable chiral models
in general: this has been the main motivation behind the present work.

We begin by fixing our conventions. The target space for the chiral models to
be considered in this paper will be a simple Lie group $G$ (which is usually,
though not necessarily, assumed to be compact) with Lie algebra $\gotg\,$, and
we shall use the trace ${\rm tr}$ in some irreducible representation to define
a) the invariant scalar product $(.\,,.)$ on $\gotg\,$, normalized so that the
long roots have length $\sqrt{\,2}$, and b) the invariant closed three-form
$\omega$ on $\gotg\,$ giving rise to the Wess-Zumino term. Explicitly, for
$\, X,Y,Z \smin \gotg\,$,
\begin{equation} \label{eq:SP}
 (X,Y)~~=~~- \, {\rm tr}\,(XY)~~~,
\end{equation}
while
\begin{equation} \label{eq:OM1}
 \omega(X,Y,Z)~~=~~{1\over 4\pi} \; {\rm tr}\,(X[Y,Z])~~~.
\end{equation}
Obviously, $(.\,,.)$ and $\omega$ extend to a biinvariant metric $(.\,,.)$
on $G$ and to a biinvariant three-form $\omega$ on $G$, respectively: the
latter can alternatively be represented in terms of the left invariant
Maurer-Cartan form $\, g^{-1} dg \,$ or right invariant Maurer-Cartan form
$\, dg \, g^{-1} \,$ on $G$, as follows:
\begin{equation} \label{eq:OM2}
 \omega~=~{1\over 12\pi} \; {\rm tr} \, (g^{-1}dg)^3~
        =~{1\over 12\pi} \; {\rm tr} \, (dg \, g^{-1})^3~~~.
\end{equation}
(Due to the Maurer-Cartan structure equation, this representation implies
that $\omega$ is indeed a closed three-form on $G$, and the normalization
in eqns (\ref{eq:OM1}) and (\ref{eq:OM2}) is chosen so that $\omega/2\pi$
generates the third de Rham cohomology group $H^3(G,\Integer)$ of $G$
over the integers, at least when $G$ is simply connected; cf.\ ref.\
\cite[p.\ 129 and Appendix 1]{FGK}. Finally, the minus sign in eqn
(\ref{eq:SP}) is introduced to ensure positive definiteness when $G$
is compact.)

In most of what follows, we shall work in terms of (arbitrary) local
coordinates $u^i$ on $G$, representing the metric $(.\,,.)$ by its
components $g_{ij}$ and the three-form $\omega$ by its components
$\omega_{ijk}$. Then the total action is the sum
\begin{equation} \label{eq:S}
 S~=~S_{CH} + n S_{WZ}
\end{equation}
of the action for the ordinary chiral model
\begin{equation} \label{eq:SCH1}
 S_{CH}~=~{1\over 2\lambda^2} \int_\Sigma d^2 x~\eta^{\mu\nu} \,
          g_{ij}(\varphi) \, \partial_\mu \varphi^i \,
                             \partial_\nu \varphi^j~~~,
\end{equation}
with coupling constant $\lambda$, and the Wess-Zumino term
\begin{equation} \label{eq:SWZ1}
 S_{WZ}~=~{1\over 6} \int_B d^3 x~\epsilon^{\kappa\lambda\mu} \,
          \omega_{ijk}(\tilde{\varphi}) \,
          \partial_\kappa \tilde{\varphi}^i \,
          \partial_\lambda \tilde{\varphi}^j \,
          \partial_\mu \tilde{\varphi}^k
       ~=~\int_B \tilde{\varphi}^\star \omega~~~.
\end{equation}
Here, $\varphi$ and $\tilde{\varphi}$ are the basic field and the extended
field of the theory, respectively, i.e., $\varphi$ is a (smooth) map from a
fixed two-dimensional Lorentz manifold $\Sigma$ to $G$ and $\tilde{\varphi}$
is a (smooth) map from an appropriate three-dimensional manifold $B$ to $G$,
chosen so that $\Sigma$ is the boundary of $B$ and $\varphi$ is the restriction
of $\tilde{\varphi}$ to that boundary:
\[
 \Sigma~=~\partial B~~~,~~~\varphi~=~\tilde{\varphi} \vert_{\Sigma}~~~.
\]
The conformally invariant WZNW model is obtained at
$\, \lambda = \sqrt{\,4\pi/|n|} \,$, while the ordinary chiral model
can be recovered in the limit $\lambda\!\rightarrow\!0$. Note that if
$\omega$ were exact, we could write $\, \omega = d\alpha \,$ to obtain
\begin{equation} \label{eq:SWZ2}
 S_{WZ}~=~{1\over 2} \int_\Sigma d^2 x~\epsilon^{\mu\nu} \,
          \alpha_{ij}(\varphi) \,
          \partial_\mu \varphi^i \, \partial_\nu \varphi^j
       ~=~\int_\Sigma \varphi^\star \alpha~~~.
\end{equation}
But of course this is not possible globally, i.e., the $\alpha_{ij}$ appearing
in this formula are neither unique nor can they be chosen so as to become the
components of a globally well-defined two-form on $G$ with respect to the
$u^i$.
Still, calculations involving quantities that arise from local variations of
the action can be performed as if this were the case, and may lead to results
that do not depend on any artificial choices. For example, recall that in the
ordinary chiral model, the canonically conjugate momenta $\pi_i$ derived from
the action $S_{CH}$ are simply given by
\begin{equation} \label{eq:MCH}
 \pi_i~=~{1\over \lambda^2} \; g_{ij}(\varphi) \, \dot{\varphi}^j
\end{equation}
and satisfy the canonical commutation relations
\begin{eqnarray} \label{eq:CCRCH}
 \{ \varphi^i(x) , \varphi^j(y) \}~=~0 &,& \{ \pi_i(x) , \pi_j(y) \}~=~0~~~,
                                                              \nonumber \\[1mm]
 \{ \varphi^i(x) , \pi_j(y) \} &=& \delta^i_j \, \delta(x-y)~~~.
\end{eqnarray}
Similarly, in the chiral model with a Wess-Zumino term, written in the form
(\ref{eq:SWZ2}), the canonically conjugate momenta $\hat{\pi}_i$ derived
from the action $S$ are given by
\begin{equation} \label{eq:MCHWZ}
 \hat{\pi}_i~=~\pi_i \, + \, n \, \alpha_{ij}(\varphi) \, \varphi^{\prime\, j}
\end{equation}
and satisfy the canonical commutation relations
\begin{eqnarray} \label{eq:CCRCHWZ}
 \{ \varphi^i(x) , \varphi^j(y) \}~=~0 &,&
 \{ \hat{\pi}_i(x) , \hat{\pi}_j(y) \}~=~0~~~,                \nonumber \\[1mm]
 \{ \varphi^i(x) , \hat{\pi}_j(y) \} &=& \delta^i_j \, \delta(x-y)~~~.
\end{eqnarray}
Note, however, that in contrast to the $\pi_i$, the $\hat{\pi}_i$ do not behave
naturally under local coordinate transformations on $G$, so that the canonical
commutation relations (\ref{eq:CCRCHWZ}) between the $\varphi^i$ and the
$\hat{\pi}_j$ look non-covariant. This suggests to consider instead the
commutation relations between the $\varphi^i$ and the $\pi_j$, which are
covariant, but exhibit non-vanishing Poisson brackets between the $\pi_i$.
Indeed, it follows from (\ref{eq:CCRCHWZ}) that
\begin{eqnarray*}
 \{ \pi_i(x) , \pi_j(y) \}
 &=& \{ \, \hat{\pi}_i(x) \, - \, n \, \alpha_{ik}(\varphi(x)) \,
                                       \varphi^{\prime k}(x) \; , \;
           \hat{\pi}_j(y) \, - \, n \, \alpha_{jl}(\varphi(y)) \,
                                       \varphi^{\prime l}(y) \, \}      \\[1mm]
 &=& n \, \left(
     - \, \{ \, \hat{\pi}_i(x) \, , \,
             \alpha_{jl}(\varphi(y)) \, \varphi^{\prime l}(y) \, \} \right.  \\
 & & \phantom{n \left( \right. \!} \left.
     - \, \{ \, \alpha_{ik}(\varphi(x)) \, \varphi^{\prime k}(x) \, , \,
             \hat{\pi}_j(y) \, \} \right)                               \\[1mm]
 &=& n \, \left(
     + \, \partial_i \alpha_{jl}(\varphi(x)) \, \varphi^{\prime l}(x) \,
          \delta(x-y) \, - \, \alpha_{ji}(\varphi(y)) \, \delta^\prime(x-y)
          \right.                                                            \\
 & & \phantom{n \left( \right. \!} \left.
     - \, \partial_j \alpha_{ik}(\varphi(x)) \, \varphi^{\prime k}(x) \,
          \delta(x-y) \, - \, \alpha_{ij}(\varphi(x)) \, \delta^\prime(x-y)
          \right)                                                       \\[1mm]
 &=& + \, n \; (\partial_i \alpha_{jk} + \partial_j \alpha_{ki})(\varphi(x)) \,
               \varphi^{\prime k}(x) \, \delta(x-y)                   \\[0.5mm]
 & & + \, n \; \partial_k \alpha_{ij}(\varphi(x)) \,
               \varphi^{\prime k}(x) \, \delta(x-y)~~~,
\end{eqnarray*}
so in the presence of the Wess-Zumino term, the commutation relations between
the $\varphi^i$ and the $\pi_j$ read
\begin{eqnarray} \label{eq:CRCHWZ}
 \{ \varphi^i(x) , \varphi^j(y) \}~=~0 &,&
 \{ \varphi^i(x) , \pi_j(y) \}~~=~~\delta^i_j \, \delta(x-y)~~~,
                                                              \nonumber \\[1mm]
 \{ \pi_i(x) , \pi_j(y) \} &=& n \, \omega_{ijk}(\varphi(x)) \,
                                    \varphi^{\prime k}(x) \, \delta(x-y)~~~.
\end{eqnarray}
They are obviously covariant (all expressions behave naturally under local
coordinate transformations on $G$), since $\omega$ is a globally well-defined
three-form on $G$.

To derive the desired current algebra, we recall next that the model under
consideration has an obvious global invariance under the product group
$\, G_L \times G_R \,$, which acts on $G$ according to
\begin{equation}
 g \, \longrightarrow \; (g_L,g_R) \cdot g = g_L \, g \, g_R^{-1}~~~.
\end{equation}
This action of the Lie group $\, G_L \times G_R \,$ induces a representation of
the corresponding Lie algebra $\, \gotg_L \oplus \gotg_R \,$ by vector fields,
associating with each generator $\; X = (X_L,X_R) \;$ in $\, \gotg_L \oplus
\gotg_R \,$ the fundamental vector field $X_G$ on $G$ given by
\begin{equation} \label{eq:FVF}
 X_G(g)~=~X_L g - g X_R~~~~~{\rm for}~g \smin G~~~.
\end{equation}
As usual, invariance of the action leads to conserved Noether currents taking
values in $\, \gotg_L \oplus \gotg_R \,$ and denoted by $j_\mu$ for the
ordinary chiral model and by $\hat{\mbox{\it \j}}_\mu$ for the chiral model
with a Wess-Zumino term. Explicitly, we have, for $\; X = (X_L,X_R) \;$ in
$\, \gotg_L \oplus \gotg_R \,$,
\begin{equation} \label{eq:NCCH1}
 (j_\mu,X)~=~- \, {1\over \lambda^2} \, g_{ij}(\varphi) \,
                  \partial_\mu \varphi^i \, X_G^j(\varphi)~~~,
\vspace{-2mm}
\end{equation}
while
\begin{equation} \label{eq:NCCHWZ1}
 (\hat{\mbox{\it \j}}_\mu,X)~=~- \; \Big( {1\over \lambda^2} \,
       g_{ij}(\varphi) \, \partial_\mu \varphi^i \,
       + \, n \, \alpha_{ij}(\varphi) \, \epsilon_{\mu\nu} \,
       \partial\>\!^\nu \varphi^i \Big) \, X_G^j(\varphi)~~~.
\vspace{1mm}
\end{equation}
In addition, an important role is played by the scalar field $j$ introduced
in ref.\ \cite{FLS}, defined by
\begin{equation} \label{eq:SF1}
 (j, X\otimes Y)~=~{1\over \lambda^2} \, g_{ij}(\varphi) \,
                   X_G^i(\varphi) \, Y_G^j(\varphi)~~~.
\vspace{1mm}
\end{equation}
(Note the additional factors ${1\over \lambda^2}$, which were absent in ref.\
\cite{FLS}.)

The commutation relations of the Noether currents $j_\mu$ and
$\hat{\mbox{\it \j}}_\mu$ under Poisson brackets can now be computed
directly. Note again, however, that in contrast to the $j_\mu$, the
$\hat{\mbox{\it \j}}_\mu$ do not behave naturally under local coordinate
transformations on $G$, so that their commutation relations look non-covariant.
This suggests to replace them by appropriate covariant currents $J_\mu$ which,
as it turns out, can be written entirely in terms of the Noether currents
$j_\mu$ for the ordinary chiral model (the exact definition will be given
below): it is the commutation relations of these covariant currents $J_\mu$
that form the current algebra we wish to compute (or at least an important
part thereof). The most efficient way of arriving at the desired result is
therefore to calculate, as an intermediate step, the commutation relations of
the Noether currents $j_\mu$, using the commutation relations
(\ref{eq:CRCHWZ}):
this can be done along the lines of ref.\ \cite{FLS}. In fact, for the ordinary
chiral model the calculation proceeds in exactly the same manner as outlined
there, while for the chiral model with a Wess-Zumino term, the only commutation
relation that changes is the Poisson bracket of two $j_0$'s, which picks up
an additional term due to the fact that the $\pi_i$ no longer commute; thus:
\begin{eqnarray} \label{eq:j0j0}
\lefteqn{\{ \, (j_0(x),X) \, , \, (j_0(y),Y) \, \}~~
 =~~\{ \pi_i(x)\, X_G^i(\varphi(x)) \, , \, \pi_j(y)\, Y_G^j(\varphi(y)) \}}
                                             \nonumber \hspace{1.7cm} \\[2.5mm]
 &=& \pi_i(x) \left( \partial_j X_G^i(\varphi(x)) \, Y_G^j(\varphi(x))
              - X_G^j(\varphi(x)) \, \partial_j Y_G^i(\varphi(x)) \right) \,
     \delta(x-y)                                              \nonumber \\[1mm]
 & & + \, n \, \omega_{ijk}(\varphi(x)) \, \varphi^{\prime k}(x) \,
          X_G^i(\varphi(x)) \, Y_G^j(\varphi(x)) \; \delta(x-y)
                                                            \nonumber \\[2.5mm]
 &=& \pi_i(x) \, [X,Y]_G^i(\varphi(x)) \, \delta(x-y)         \nonumber \\[1mm]
 & & + \, n \; \omega(\varphi(x)) \left( \varphi^\prime(x) ,
          X_G(\varphi(x)) , Y_G(\varphi(x)) \right) \, \delta(x-y)
                                                            \nonumber \\[2.5mm]
 &=& - \, (j_0(x),[X,Y]) \, \delta(x-y)                       \nonumber \\[1mm]
 & & + \, n \; \omega(\varphi(x)) \left( \varphi^\prime(x) ,
          X_L \varphi(x) - \varphi(x) X_R ,
          Y_L \varphi(x) - \varphi(x) Y_R \right) \, \delta(x-y)~~~~
                                                              \nonumber \\[2mm]
 & &
\end{eqnarray}
For later reference, we also list the commutation relations which have remained
unchanged:
\begin{eqnarray}
\lefteqn{\{ \, (j_0(x),X) \, , \, (j_1(y),Y) \, \}}
                                               \nonumber \hspace{1.7cm} \\[1mm]
 &=& - \, (j_1(x),[X,Y]) \, \delta(x-y) \;
     + \; (j(y), X\otimes Y) \, \delta^\prime(x-y)~~~,~~~~
                                                        \label{eq:j0j1} \\[2mm]
\lefteqn{\{ \, (j_1(x),X) \, , \, (j_1(y),Y) \, \}~~=~~0~~~,~~~~}
                                         \hspace{1.7cm} \label{eq:j1j1} \\[3mm]
\lefteqn{\{ \, (j_0(x),X) \, , \, (j(y), Y\otimes Z) \, \}}
                                               \nonumber \hspace{1.7cm} \\[1mm]
 &=& - \, ( \, j(x) \, , \, [X,Y]\otimes Z + Y\otimes [X,Z] \, ) \;
     \delta(x-y)~~~,~~~~                                 \label{eq:j0j} \\[2mm]
\lefteqn{\{ \, (j_1(x),X) \, , \, (j(y), Y\otimes Z) \, \}~~=~~0~~~.~~~~}
                                          \hspace{1.7cm} \label{eq:j1j}
\end{eqnarray}
(The additional factors ${1\over \lambda^2}$ drop out completely: they have
been absorbed into the normalizations of the $j_\mu$ and $j$.)

Before proceeding further, we find it convenient to pass to more
standard notation, writing $g$ and $\tilde{g}$, rather than $\varphi$ and
$\tilde{\varphi}$, for the basic field and the extended field of the theory,
respectively, and using the explicit definitions (\ref{eq:SP}) of the metric
$(.\,,.)$ on $G$ and (\ref{eq:OM1}) of the three-form $\omega$ on $G$. Then
\begin{equation} \label{eq:SCH2}
 S_{CH}~=~- \, {1\over 2\lambda^2} \int d^2 x~\eta^{\mu\nu} \; {\rm tr}
               \left( g^{-1} \partial_\mu g \, g^{-1} \partial_\nu g
\right)~~~,
\vspace{-1mm}
\end{equation}
while
\vspace{1mm}
\begin{equation} \label{eq:SWZ3}
 S_{WZ}~=~{1\over 4\pi} \int_0^1 dr \int d^2 x~\epsilon^{\mu\nu} \; {\rm tr}
          \left( \tilde{g}^{-1} \partial_r \tilde{g} \,
                 \tilde{g}^{-1} \partial_\mu \tilde{g} \,
                 \tilde{g}^{-1} \partial_\nu \tilde{g} \right)~~~.
\vspace{2mm}
\end{equation}
(Here, the extended field $\tilde{g}$ is assumed to be constant outside a
tubular neighborhood $\, \Sigma \times [0,1] \,$ of the boundary $\Sigma$
of $B$, and $r$ is the coordinate normal to the boundary.) Next, we decompose
the currents $j_\mu$ and $J_\mu$, both of which take values in $\, \gotg_L
\oplus \gotg_R \,$, into left and right currents, all of which take values in
$\gotg\,$: $\; j_\mu = (j_\mu^L,j_\mu^R) \, , \; J_\mu = (J_\mu^L,J_\mu^R) \,$.
Explicitly,
\begin{eqnarray}
 j_\mu^L &=& - \, {1\over \lambda^2} \, \partial_\mu g \, g^{-1}~~~,
                                                             \label{eq:NC2a} \\
 j_\mu^R &=& + \, {1\over \lambda^2} \, g^{-1} \partial_\mu g \,~~~,
                                                             \label{eq:NC2b}
\end{eqnarray}
and, by definition,
\begin{eqnarray}
 J_\mu^L &=& \left( \eta_{\mu\nu} + {n\lambda^2 \over 4\pi} \,
                    \epsilon_{\mu\nu} \right) j^{L\,\nu}~~
         =~~ - \, {1\over \lambda^2} \,
                   \left( \eta_{\mu\nu} + {n\lambda^2 \over 4\pi} \,
                    \epsilon_{\mu\nu} \right) \partial\,^\nu g \, g^{-1}~~~,
                                                             \label{eq:NC3a} \\
 J_\mu^R &=& \left( \eta_{\mu\nu} - {n\lambda^2 \over 4\pi} \,
                    \epsilon_{\mu\nu} \right) j^{R\,\nu}~~
         =~~ + \, {1\over \lambda^2} \,
                   \left( \eta_{\mu\nu} - {n\lambda^2 \over 4\pi} \,
                    \epsilon_{\mu\nu} \right) g^{-1} \partial\,^\nu g \,~~~,
                                                             \label{eq:NC3b}
\end{eqnarray}
while the scalar field $j$, when viewed as taking values in the space
of endomorphisms of $\, \gotg_L \oplus \gotg_R \,$, is given by the
$(2\times 2)$-block matrix
\begin{equation} \label{eq:SF2}
 j~=~{1\over \lambda^2} \left( \begin{array}{cc}
                                1 & - {\rm Ad}(g) \\
                                - {\rm Ad}(g)^{-1} & 1
                               \end{array} \right)~~~.
\end{equation}
In other words, for $\; X = (X_L,X_R) \;$ in $\, \gotg_L \oplus \gotg_R \,$,
\begin{equation} \label{eq:SF3}
 j(X)~=~{1\over \lambda^2} \bigg(
        X_L - {\rm Ad}(g) X_R \, , \, X_R - {\rm Ad}(g)^{-1} X_L \bigg)~~~.
\end{equation}
It can be shown that the covariant currents $J_\mu$ defined by
eqns (\ref{eq:NC3a},\ref{eq:NC3b}) differ from the Noether currents
$\hat{\mbox{\it \j}}_\mu$ for the chiral model with a Wess-Zumino term
by a total curl, and that current conservation (which for both types of
currents has the same physical content, because a total curl is automatically
conserved) is identical with the equations of motion of the theory; cf., e.g.,
ref.\ \cite{FZ} for more details\footnote{Our conventions here differ from
those of ref.\ \cite{FZ} in that we write $\lambda^2$ instead of $\lambda$
and use \linebreak $\; \epsilon^{01} = + 1 \, , \epsilon_{01} = - 1 \,$:
thus the formulas valid for $n>0$ ($n<0$) there correspond to the formulas
valid for $n<0$ ($n>0$) here.}. (Note also that the additional factors of
${1\over 2}$ in eqns (24) and (25) of ref.\ \cite{FLS}, as compared to
eqns (\ref{eq:NC2a},\ref{eq:NC2b}) and (\ref{eq:SF2}) above, are due to the
identification, performed there, of $G$ with the Riemannian symmetric space
$\; G \times G / \Delta G \,$, which results in $G$ being equipped with a
biinvariant metric that is twice the one used here.)

Now in terms of an arbitrary basis $(T_a)$ of $\gotg\,$, with structure
constants $f_{ab}^c$ defined by \mbox{$\; [T_a,T_b] = f_{ab}^c T_c \;$},
the various currents are represented by their components
\begin{eqnarray}
 j_{\mu,a}^L &=~~(j_\mu,T_a^L)~~=& - \, {\rm tr} \, (j_\mu^L \, T_a)~~~,
                                                             \label{eq:NCCL} \\
 j_{\mu,a}^R &=~~(j_\mu,T_a^R)~~=& - \, {\rm tr} \, (j_\mu^R \, T_a)~~~,
                                                             \label{eq:NCCR} \\
 J_{\mu,a}^L &=~~(J_\mu,T_a^L)~~=& - \, {\rm tr} \, (J_\mu^L \, T_a)~~~,
                                                             \label{eq:CCCL} \\
 J_{\mu,a}^R &=~~(J_\mu,T_a^R)~~=& - \, {\rm tr} \, (J_\mu^R \, T_a)~~~,
                                                             \label{eq:CCCR}
\end{eqnarray}
and the scalar field $j$ by its components
\begin{equation} \label{eq:SFCDIAG}
 \eta_{ab}~=~(j, T_a^L\otimes T_b^L)~=~(j, T_a^R\otimes T_b^R)~
 =~- \, {1\over \lambda^2} \, {\rm tr} \left( T_a T_b \right)~~~,
\end{equation}
\begin{equation} \label{eq:SFCOFFDIAG}
 t_{ab}~=~(j, T_a^L\otimes T_b^R)~
 =~{1\over \lambda^2} \, {\rm tr} \left( g^{-1} \, T_a \, g \, T_b \right)~~~,
\vspace{-1mm}
\end{equation}
where
\vspace{1mm}
\begin{equation} \label{eq:TLTR}
 T_a^L~=~(T_a,0)~~~,~~~T_a^R~=~(0,T_a)~~~.
\vspace{2mm}
\end{equation}
With this notation, eqns (\ref{eq:j0j0})--(\ref{eq:j1j1}) can easily be seen to
imply the following commutation relations for the components of the currents
$j_\mu$:
\pagebreak
\begin{eqnarray}
 \{ j_{0,a}^L(x) , j_{0,b}^L(y) \}
 &=& - \, f_{ab}^c \, j_{0,c}^L(x) \, \delta(x-y) \,
     + \, {n\lambda^2 \over 4\pi} \, f_{ab}^c \, j_{1,c}^L(x) \,
     \delta(x-y)~~~,~~~~                              \label{eq:CAL0L0} \\[1mm]
 \{ j_{0,a}^L(x) , j_{1,b}^L(y) \}
 &=& - \, f_{ab}^c \, j_{1,c}^L(x) \, \delta(x-y) \,
     + \, \eta_{ab} \, \delta^\prime(x-y)~~~,~~~~     \label{eq:CAL0L1} \\[2mm]
 \{ j_{1,a}^L(x) , j_{1,b}^L(y) \} &=& \, 0~~~,~~~~   \label{eq:CAL1L1} \\[2mm]
 \{ j_{0,a}^R(x) , j_{0,b}^R(y) \}
 &=& - \, f_{ab}^c \, j_{0,c}^R(x) \, \delta(x-y) \,
     - \, {n\lambda^2 \over 4\pi} \, f_{ab}^c \, j_{1,c}^R(x) \,
     \delta(x-y)~~~,~~~~                              \label{eq:CAR0R0} \\[1mm]
 \{ j_{0,a}^R(x) , j_{1,b}^R(y) \}
 &=& - \, f_{ab}^c \, j_{1,c}^R(x) \, \delta(x-y) \,
     + \, \eta_{ab} \, \delta^\prime(x-y)~~~,~~~~     \label{eq:CAR0R1} \\[2mm]
 \{ j_{1,a}^R(x) , j_{1,b}^R(y) \} &=& \, 0~~~,~~~~   \label{eq:CAR1R1} \\[2mm]
 \{ j_{0,a}^L(x) , j_{0,b}^R(y) \}
 &=& {n\lambda^2 \over 4\pi} \, t_{ab}^\prime(x) \, \delta(x-y)~~~,~~~~
                                                      \label{eq:CAL0R0} \\[1mm]
 \{ j_{0,a}^L(x) , j_{1,b}^R(y) \}
 &=& t_{ab}(y) \, \delta^\prime(x-y)~~~,~~~~          \label{eq:CAL0R1} \\[2mm]
 \{ j_{0,a}^R(x) , j_{1,b}^L(y) \}
 &=& t_{ba}(y) \, \delta^\prime(x-y)~~~,~~~~          \label{eq:CAR0L1} \\[2mm]
 \{ j_{1,a}^L(x) , j_{1,b}^R(y) \} &=& \, 0~~~.~~~~   \label{eq:CAL1R1}
\end{eqnarray}
They must be supplemented by the commutation relations between the components
of the currents $j_\mu$ and those of the field $t$, which follow from eqns
(\ref{eq:j0j}) and (\ref{eq:j1j}):
\begin{eqnarray}
 \{ j_{0,a}^L(x) , t_{bc}(y) \}
 &=& - \, f_{ab}^d \, t_{dc}(x) \, \delta(x-y)~~~,~~~~ \label{eq:CAL0T} \\[1mm]
 \{ j_{0,a}^R(x) , t_{bc}(y) \}
 &=& - \, f_{ac}^d \, t_{bd}(x) \, \delta(x-y)~~~,~~~~ \label{eq:CAR0T} \\[2mm]
 \{ j_{1,a}^L(x) , t_{bc}(y) \} &=& \, 0~~~,~~~~       \label{eq:CAL1T} \\[1mm]
 \{ j_{1,a}^R(x) , t_{bc}(y) \} &=& \, 0~~~,~~~~       \label{eq:CAR1T}
\end{eqnarray}
Finally, the components of the field $t$ commute among themselves:
\begin{equation}
 \{ t_{ab}(x) , t_{cd}(y) \}~~=~~0~~~.~~~~                      \label{eq:CATT}
\end{equation}
In terms of light-cone components, which for any vector $a_\mu$ are defined
by $~a_\pm = a_0 \pm a_1 \,$, these commutation relations take the form
\begin{eqnarray}
 \{ j_{\pm,a}^L(x) , j_{\pm,b}^L(y) \}
 &=& - \, {1\over 2} \, f_{ab}^c \,
     \Big[ \Big( 3 \mp {n\lambda^2 \over 4\pi} \Big) \, j_{\pm,c}^L(x) \, -
           \Big( 1 \mp {n\lambda^2 \over 4\pi} \Big) \, j_{\mp,c}^L(x) \, \Big]
     \, \delta(x-y)~~~~                                       \nonumber \\[2mm]
 & & \pm \, 2 \, \eta_{ab} \, \delta^\prime(x-y)~~~,\label{eq:CALPMLPM1}\\[3mm]
 \{ j_{\pm,a}^R(x) , j_{\pm,b}^R(y) \}
 &=& - \, {1\over 2} \, f_{ab}^c \,
     \Big[ \Big( 3 \pm {n\lambda^2 \over 4\pi} \Big) \, j_{\pm,c}^R(x) \, -
           \Big( 1 \pm {n\lambda^2 \over 4\pi} \Big) \, j_{\mp,c}^R(x) \, \Big]
     \, \delta(x-y)~~~~                                       \nonumber \\[2mm]
 & & \pm \, 2 \, \eta_{ab} \, \delta^\prime(x-y)~~~,\label{eq:CARPMRPM1}\\[3mm]
 \{ j_{\pm,a}^L(x) , j_{\pm,b}^R(y) \}
 &=& \pm\, \Big[ \Big( 1 \mp {n\lambda^2 \over 4\pi} \Big) \, t_{ab}(x) \, + \,
                 \Big( 1 \pm {n\lambda^2 \over 4\pi} \Big) \, t_{ab}(y) \,
\Big]
     \, \delta^\prime(x-y)~,~~~~~~                  \label{eq:CALPMRPM1}\\[4mm]
 \{ j_{\pm,a}^L(x) , j_{\mp,b}^L(y) \}
 &=& - \, {1\over 2} \, f_{ab}^c \,
       \Big[ \Big( 1 - {n\lambda^2 \over 4\pi} \Big) \, j_{+,c}^L(x) \, +
             \Big( 1 + {n\lambda^2 \over 4\pi} \Big) \, j_{-,c}^L(x) \, \Big]
     \, \delta(x-y)~,~~~~~~                         \label{eq:CALPMLMP1}\\[2mm]
 \{ j_{\pm,a}^R(x) , j_{\mp,b}^R(y) \}
 &=& - \, {1\over 2} \, f_{ab}^c \,
       \Big[ \Big( 1 + {n\lambda^2 \over 4\pi} \Big) \, j_{+,c}^R(x) \, +
             \Big( 1 - {n\lambda^2 \over 4\pi} \Big) \, j_{-,c}^R(x) \, \Big]
     \, \delta(x-y)~,~~~~~~                         \label{eq:CARPMRMP1}\\[3mm]
 \{ j_{\pm,a}^L(x) , j_{\mp,b}^R(y) \}
 &=& \mp \, \Big( 1 \mp {n\lambda^2 \over 4\pi} \Big) \, t^\prime_{ab}(x) \,
     \delta(x-y)~,~~~~~~                            \label{eq:CALPMRMP1}\\[2mm]
 \{ j_{\pm,a}^R(x) , j_{\mp,b}^L(y) \}
 &=& \mp \, \Big( 1 \pm {n\lambda^2 \over 4\pi} \Big) \, t^\prime_{ba}(x) \,
     \delta(x-y)~,~~~~~~                            \label{eq:CARPMLMP1}
\vspace{2mm}
\end{eqnarray}
and
\begin{eqnarray}
 \{ j_{\pm,a}^L(x) , t_{bc}(y) \}
 &=& - \, f_{ab}^d \, t_{dc}(x) \, \delta(x-y)~~~,~~~~\label{eq:CALPMT1}\\[1mm]
 \{ j_{\pm,a}^R(x) , t_{bc}(y) \}
 &=& - \, f_{ac}^d \, t_{bd}(x) \, \delta(x-y)~~~,~~~~\label{eq:CARPMT1}
\vspace{2mm}
\end{eqnarray}
together with eqn (\ref{eq:CATT}).

Passing to the commutation relations for the currents $J_\mu$ instead
of the currents $j_\mu$ is now trivial, because according to eqns
(\ref{eq:NC3a},\ref{eq:NC3b}), their light-cone components just differ
by a normalization factor:
\begin{eqnarray}
 J_\pm^L &=& \Big( 1 \pm {n\lambda^2 \over 4\pi} \Big) \, j_\pm^L~~~,~~~~    \\
 J_\pm^R &=& \Big( 1 \mp {n\lambda^2 \over 4\pi} \Big) \, j_\pm^L~~~,~~~~
\end{eqnarray}
We nevertheless write down the corresponding formulas, because they constitute
the main result of this paper; they read
\vspace{2mm}
\begin{eqnarray}
\lefteqn{\{ J_{\pm,a}^L(x) , J_{\pm,b}^L(y) \}} \nonumber \hspace{1.1cm}\\[2mm]
 &=& - \, {1\over 2} \, \Big( 1 \pm {n\lambda^2 \over 4\pi} \Big) \, f_{ab}^c
\,
     \Big[ \Big( 3 \mp {n\lambda^2 \over 4\pi} \Big) \, J_{\pm,c}^L(x) \, - \,
           \Big( 1 \pm {n\lambda^2 \over 4\pi} \Big) \, J_{\mp,c}^L(x) \, \Big]
     \, \delta(x-y)                                           \nonumber \\[1mm]
 & & \pm \, 2 \, \Big( 1 \pm {n\lambda^2 \over 4\pi} \Big)^2 \,
          \eta_{ab} \, \delta^\prime(x-y)~,         \label{eq:CALPMLPM2}\\[4mm]
\lefteqn{\{ J_{\pm,a}^R(x) , J_{\pm,b}^R(y) \}} \nonumber \hspace{1.1cm}\\[2mm]
 &=& - \, {1\over 2} \, \Big( 1 \mp {n\lambda^2 \over 4\pi} \Big) \, f_{ab}^c
\,
     \Big[ \Big( 3 \pm {n\lambda^2 \over 4\pi} \Big) \, J_{\pm,c}^R(x) \, - \,
           \Big( 1 \mp {n\lambda^2 \over 4\pi} \Big) \, J_{\mp,c}^R(x) \, \Big]
     \, \delta(x-y)                                           \nonumber \\[1mm]
 & & \pm \, 2 \, \Big( 1 \mp {n\lambda^2 \over 4\pi} \Big)^2 \,
          \eta_{ab} \, \delta^\prime(x-y)~,         \label{eq:CARPMRPM2}
\end{eqnarray}
\pagebreak
\begin{eqnarray}
\lefteqn{\{ J_{\pm,a}^L(x) , J_{\pm,b}^R(y) \}} \nonumber \hspace{1.1cm}\\[1mm]
 &=& \pm \, \Big( 1 - \Big( {n\lambda^2 \over 4\pi} \Big)^2 \Big) \,
     \Big[ \Big( 1 \mp {n\lambda^2 \over 4\pi} \Big) \, t_{ab}(x) \, + \,
           \Big( 1 \pm {n\lambda^2 \over 4\pi} \Big) \, t_{ab}(y) \, \Big] \,
     \delta^\prime(x-y)~,~~~~~~                     \label{eq:CALPMRPM2}\\[4mm]
\lefteqn{\{ J_{\pm,a}^L(x) , J_{\mp,b}^L(y) \}} \nonumber \hspace{1.1cm}\\[1mm]
 &=& - \, {1\over 2} \, f_{ab}^c \,
     \Big[ \Big( 1 - {n\lambda^2 \over 4\pi} \Big)^2 \, J_{+,c}^L(x) \, + \,
           \Big( 1 + {n\lambda^2 \over 4\pi} \Big)^2 \, J_{-,c}^L(x) \, \Big]
     \, \delta(x-y)~,~~~~~~                         \label{eq:CALPMLMP2}\\[2mm]
\lefteqn{\{ J_{\pm,a}^R(x) , J_{\mp,b}^R(y) \}} \nonumber \hspace{1.1cm}\\[1mm]
 &=& - \, {1\over 2} \, f_{ab}^c \,
     \Big[ \Big( 1 + {n\lambda^2 \over 4\pi} \Big)^2 \, J_{+,c}^R(x) \, + \,
           \Big( 1 - {n\lambda^2 \over 4\pi} \Big)^2 \, J_{-,c}^R(x) \, \Big]
     \, \delta(x-y)~,~~~~~~                         \label{eq:CARPMRMP2}\\[7mm]
\lefteqn{\{ J_{\pm,a}^L(x) , J_{\mp,b}^R(y) \}~~
 =~~\mp \, \Big( 1 - \Big( {n\lambda^2 \over 4\pi} \Big)^2 \Big)
           \Big( 1 \pm {n\lambda^2 \over 4\pi} \Big) \,
    t^\prime_{ab}(x) \, \delta(x-y)~,}              \label{eq:CALPMRMP2}\\[2mm]
\lefteqn{\{ J_{\pm,a}^R(x) , J_{\mp,b}^L(y) \}~~
 =~~\mp \, \Big( 1 - \Big( {n\lambda^2 \over 4\pi} \Big)^2 \Big)
           \Big( 1 \mp {n\lambda^2 \over 4\pi} \Big) \,
    t^\prime_{ba}(x) \, \delta(x-y)~,}              \label{eq:CARPMLMP2}
\vspace{3mm}
\end{eqnarray}
and
\begin{eqnarray}
 \{ J_{\pm,a}^L(x) , t_{bc}(y) \}
 &=& - \, \Big( 1 \pm {n\lambda^2 \over 4\pi} \Big) \, f_{ab}^d \,
       t_{dc}(x) \, \delta(x-y)~~~,~~~~               \label{eq:CALPMT2}\\[1mm]
 \{ J_{\pm,a}^R(x) , t_{bc}(y) \}
 &=& - \, \Big( 1 \mp {n\lambda^2 \over 4\pi} \Big) \, f_{ac}^d \,
       t_{bd}(x) \, \delta(x-y)~~~,~~~~               \label{eq:CARPMT2}
\vspace{3mm}
\end{eqnarray}
together with eqn (\ref{eq:CATT}).

Note that all these Poisson bracket relations are consistent with the discrete
symmetries of the model, which are built from
\vspace{-4mm}
\begin{itemize}
 \item the exchange $\, L\leftrightarrow R \,$ of left and right components,
       which originates from the transformation $\, g \rightarrow g^{-1} \,$
       in $G$ and must therefore be accompanied by a transposition of indices
       in $t$,
\vspace{-1mm}
 \item the exchange $\, +\leftrightarrow - \,$ of light-cone components,
       which originates from a parity transformation and must therefore
       be accompanied by an extra change of sign in spatial derivatives
       of $t$ and of the delta function,
\vspace{-1mm}
 \item the exchange $\, n\leftrightarrow -n$.
\vspace{-4mm}
\end{itemize}
More specifically, the action of the model and the Poisson bracket relations
are invariant under the combination of any two of these three symmetries:
\vspace{-4mm}
\begin{tabbing}
a) \= $+\leftrightarrow -$ \= and \= $n\leftrightarrow -n$, \kill
a) \= $L\leftrightarrow R$ \> and \> $n\leftrightarrow -n$, \\[2mm]
b) \= $+\leftrightarrow -$ \> and \> $n\leftrightarrow -n$, \\[2mm]
c) \= $L\leftrightarrow R$ \> and \> $+\leftrightarrow -$.
\vspace{-5mm}
\end{tabbing}
Hence the case $n\!<\!0$ can be reduced to the case $n\!>\!0$ by exchanging
either left and right components or light-cone components, so we may assume
without loss of generality that $n\!>\!0$. Next, recall that the critical
value of the coupling constant $\lambda$ is $\, \lambda = \sqrt{\,4\pi/n} \,$;
here
\begin{eqnarray}
 J_+^L~=~2 \, j_+^L~&,&~J_-^L~=~0~~~,                                        \\
 J_-^R~=~2 \, j_-^R~&,&~J_+^R~=~0~~~.
\end{eqnarray}
The vanishing of the two current components $J_-^L$ and $J_+^R$ can be
interpreted as constraints, and inspection of the commutation relations
(\ref{eq:CALPMLPM2})--(\ref{eq:CARPMT2}) (for $\, \lambda^2 = 4\pi/n$)
shows that these constraints are first class, whereas the remaining
commutation relations state that the non-vanishing current components
$J_+^L$ and $J_-^R$ satisfy a Kac-Moody algebra, differing only in the
sign of the central extension, and commute with each other:
\begin{eqnarray}
 \{ J_{+,a}^L(x) , J_{+,b}^L(y) \}
 &=& - \, 2 \, f_{ab}^c \, J_{+,c}^L(x) \, \delta(x-y) \,
     + \, 8 \, \eta_{ab} \, \delta^\prime(x-y)~~~,                           \\
 \{ J_{-,a}^R(x) , J_{-,b}^R(y) \}
 &=& - \, 2 \, f_{ab}^c \, J_{-,c}^R(x) \, \delta(x-y) \,
     - \, 8 \, \eta_{ab} \, \delta^\prime(x-y)~~~,                           \\
 \{ J_{+,a}^L(x) , J_{-,b}^R(y) \} &=& 0~~~.
\end{eqnarray}
On the other hand, by taking the limit $\lambda\!\rightarrow\!0$, or simply
by setting $n\!=\!0$, we recover the current algebra for the ordinary chiral
model derived in ref.\ \cite{FLS} from eqns (\ref{eq:CAL0L0})--(\ref{eq:CAR1T})
(which are equivalent to eqns (\ref{eq:CALPMLPM1})--(\ref{eq:CARPMT1}) or
(\ref{eq:CALPMLPM2})--(\ref{eq:CARPMT2})), together with eqn (\ref{eq:CATT}).




\begin{thebibliography}{9}

\bibitem{W} E.~Witten: {\em Non-Abelian Bosonization in Two Dimensions},
 Commun.\ Math.\ Phys.\ {\bf 92} (1984) 455-472.

\bibitem{FaTa} L.D.~Faddeev and L.A.~Takhtajan: {\em Hamiltonian Methods in the
 Theory of Solitons}, Springer-Verlag, Berlin 1987.

\bibitem{FLS}
 M.~Forger, J.~Laartz and U.~Sch\"aper: {\em Current Algebra of Classical
 Non-Linear Sigma Models}, Freiburg University preprint THEP 91/10, to be
 published in Commun.\ Math.\ Phys..

\bibitem{FGK}
 G.~Felder, K.~Gaw\c{e}dzki and A.~Kupiainen: {\em Spectra of
 Wess-Zumino-Witten Models with Arbitrary Simple Groups},
 Commun.\ Math.\ Phys.\ {\bf 117} (1988) 127-158.

\bibitem{FZ}
 M.~Forger and P.~Zizzi: {\em Twisted Chiral Models with Wess-Zumino Terms,
 and Strings}, Nucl.\ Phys.\ {\bf B287} (1987) 131-143.

\bibitem{AR}
 E.~Abdalla, K.D.~Rothe, Phys.\ Rev.\ {\bf D} (1989).

\bibitem{AAR}
 E.~Abdalla, M.C.B.~Abdalla and K.D.~Rothe: {\em Non perturbative methods in
 two-dimensional quantum field theory}, World Scientific, 1991.

\end{thebibliography}
\end{document}